\documentclass[useAMS,usenatbib, twocolumn]{mn2e}

\usepackage{graphicx}

\newcommand{\gtrsim}{\,\rlap{\lower3.7pt\hbox{$\mathchar\sim$}}
\raise1pt\hbox{$>$}\,}
\newcommand{\lesssim}{\,\rlap{\lower3.7pt\hbox{$\mathchar\sim$}}
\raise1pt\hbox{$<$}\,}

\newcommand{\nc}{\newcommand}

\nc{\be}[1]{\begin{equation}\mbox{$\label{#1}$}}
\nc{\bea}[1]{\begin{eqnarray} \mbox{$\label{#1}$}}
\nc{\Section}[2]{\section{#2}\label{#1}}
\nc{\Bibitem}[1]{\bibitem{#1}}
\nc{\Label}[1]{\label{#1}}

\nc{\eea}{\end{eqnarray}}
\nc{\ee}{\end{equation}}

\nc{\Mnu}{M_\nu}
\nc{\bee}{\begin{equation}}
\nc{\ene}{\end{equation}}
\nc{\hdp}{\sigma_8 (\Omega_{\rm m}/0.3)^{0.37}}
\nc{\avis}{\alpha_{vis}}
\nc{\cvis}{c^2_{vis}}
\nc{\clam}{c^2_{lam}}

\begin{document}

\title{Constraining Dark Energy Anisotropic Stress}
\author[Mota, Kristiansen, Koivisto, Groeneboom]{David F. Mota$^1$, Jostein R. Kristiansen$^2$, Tomi Koivisto$^{3,4}$,  Nicolaas E. Groeneboom$^2$\\
$^1$Institute of Theoretical Physics, University of Heidelberg, 69120 Heidelberg, Germany\\
$^2$Institute of Theoretical Astrophysics, University of Oslo, Box 1029, 0315 Oslo, Norway\\
$^3$Helsinki Institute of Physics, P.O. Box 64, FIN-00014 Helsinki, Finland\\
$^4$Department of Physical Sciences, Helsinki University, P.O. Box 64, FIN-00014 Helsinki, Finland}


\maketitle

\begin{abstract}
We investigate the possibility of using cosmological observations to probe and constrain an imperfect dark energy fluid. 
We consider a general parameterization of the dark energy component  accounting for an
equation of state, speed of sound and viscosity. We use  present and future data from the cosmic microwave background
radiation (CMB), large scale structures and supernovae type Ia. 
 We find
that  both the speed of sound and viscosity parameters are
difficult to nail down with the present cosmological data. Also, we argue that it will be hard to improve the constraints
significantly with future CMB data sets.
The implication is that a perfect fluid description might ultimately turn out to be a phenomenologically sufficient description of all 
the observational consequences of dark energy. The fundamental lesson is however that even then one cannot exclude, by appealing to observational 
evidence alone, the possibility of imperfectness in dark energy. 

\end{abstract}


\maketitle

\section{Introduction}   

The observed accelerated  expansion of the universe (see e.g.\citep{astier:2006}) is usually ascribed to the existence of a cosmic fluid with a 
negative pressure comparable to its small but cosmologically significant energy density. While numerous possible origins have been proposed for this 
dark energy (see \citep{Copeland:2006wr} for a review) most of them are variations of the cosmological constant or
the scalar field scenario \citep{Wetterich:1987fm}. These include several different kinds of couplings to 
matter (baryons, neutrinos, cold dark matter\citep{Amendola:1999er,brook,mota,mota2,Koivisto:2005nr}), couplings to gravity 
(the curvature scalar, Lovelock invariants, etc \citep{Koivisto:2006xf,Koivisto:2006ai}), 
and also non-canonical kinetic terms (phantoms, tachyons, K-essence, etc \citep{Gibbons:2002md,ArmendarizPicon:2000ah}). 
With such a zoo of models, it is important to investigate,  and search for,  some particular feature or property of  dark energy, which could be  
used to rule out some of (or at least distinguish among) all these possibilities.

It is already well known in the literature that to discriminate between different classes of models, it is not sufficient to consider just the background expansion. 
One has also to study the evolution of
cosmological perturbations. General Relativity dictates that any component with its equation of state $w \neq -1$, where $w=-1$ corresponds to the 
cosmological constant, must fluctuate. Therefore a generic dark energy component has perturbations which couple to matter perturbations. These however can 
be small, since if one has $w \approx -1$, the component can be nearly smooth, and if the Jeans length of dark energy is large, its perturbations may 
be confined to very large scales only. This is typical for the usual minimally coupled quintessence models, since their sound speed of perturbations, $\clam$, is 
 equal to the light speed, $\clam=1$, which sets a large Jeans length \citep{Bean:2003fb,xia}. However,  for some  other  dark energy candidates that is not the case 
\citep{Mota:2004pa,Bagla:2002yn,Bento:2002ps}, and such feature might help to differentiate between these classes of models. 

In addition to the $w$ and $\clam$, there is an 
important characteristic of a general cosmic fluid which is its anisotropic stress $\sigma$ \citep{Hu:1998kj}. This vanishes for a 
minimally coupled scalar field and perfect fluid candidates, but is a generic property of  realistic fluids with finite shear viscous coefficients 
\citep{Schimd:2006pa,Brevik:2004sd,Nojiri:2005sr}.  Basically, while
$w$ and $\clam$ determine respectively the background and perturbative
pressure of the fluid that is rotationally invariant, $\sigma$ quantifies
how much the pressure of the fluid varies with direction. In fact the anisotropic stress 
perturbation is crucial to the understanding of evolution of 
inhomogeneities in the early, radiation dominated universe \citep{Hu:1998kj,koivisto:2005}.  
Therefore an obviously interesting question is whether present 
observational data could allow for an anisotropic stress perturbation in the 
late universe which is dominated by the mysterious dark energy fluid \citep{ichiki:2007,barrow}. 

The cosmological effects of 
$\sigma$ due to possible viscosity of dark energy are however quite neglected in the literature.
The main reason for disregarding the anisotropic stress in the dark energy 
fluid might be that conventional dark energy candidates, such as the 
cosmological constant or canonical scalar fields,  
are perfect fluids with $\sigma=0$.  However, since there is no fundamental theoretical model to 
describe dark energy, there are no strong reasons to stick to such assumption. 
Moreover, coupled scalar fields have indeed a non-negligible anisotropic stress\citep{Schimd:2006pa}, and  dark energy vector field candidates (which have been proposed
in  \citep{Armendariz-Picon:2004pm,Kiselev:2004py,Zimdahl:2000zm}) also have $\sigma\neq 0$.

The aim of the present paper is to investigate the potential cosmological signatures of a very general dark energy component, which is characterize by 
an equation of state, sound speed and anisotropic stress.  In section \ref{para} we review the 
parameterization of a generalized cosmological fluid and comment its relation to some recent studies of anisotropies dark energy. The parameterization 
will then be subjected to the most detailed and most extensive scrutiny this far. The data and method utilized for this are described in section
\ref{data}. In the section \ref{cons1} we use the most recent cosmological data to constrain the properties of dark energy. Section
\ref{cons2} is devoted to investigate how much the future data could be able to improve the constraints. We conclude by stating the fundamental 
uncertainty in the properties of dark energy but also mention some cases where a positive detection could be established.  

\section{Dark energy stress parameterization}
\label{para}

Consider a general fluid with the energy momentum tensor
\be{fluid}
T_{\mu\nu}= \rho u_\mu u_\nu + ph_{\mu\nu} + \Sigma_{\mu\nu},
\ee
where $u_\mu$ is the four-velocity of the fluid, and the projection
tensor $h_{\mu\nu}$ is defined as $h_{\mu\nu} \equiv g_{\mu\nu} + u_\mu u_\nu$.
Here $\Sigma_{\mu\nu}$ can include only spatial inhomogeneity. At the background level, the evolution of the fluid is determined by
the continuity equation,
\be{continuity}
\dot{\rho} + 3H(1+w)\rho = 0.
\ee
The effects to the overall expansion are therefore determined by the equation of state $w$ alone. 
We define a {\it perfect} fluid by the condition $\Sigma_{\mu\nu}=0$. The condition for the {\it adiabaticity} of a fluid is 
$p=p(\rho)$, which implies that the evolution of the sound speed is determined by the equation of state alone. Generally, however, the sound 
speed is defined as the ratio of pressure and density perturbations in the 
frame comoving with the dark energy fluid \citep{Weller:2003hw},
\be{cs}
\clam \equiv \frac{\delta p}{\delta \rho}_{|{de}}.
\ee
In the adiabatic situation one has $\clam = d p/d\rho = \frac{\dot{p}}{\dot{\rho}} = w - \frac{\dot{w}}{3H(1+w)}$, but
in general the sound speed is an independent variable. In the following we will consider a constant equation of state  for simplicity.
 
Taking these considerations into account, the evolution equations for the
dark energy density perturbation $\delta$ and velocity potential
$\theta$ in the synchronous gauge \citep{Ma:1995ey}, can be written as
\bea{deltaevol}
\dot{\delta}    &=& 
-(1+w)\left\{\left[k^2+9H^2(\clam-w)\right]\frac{\theta}{k^2} 
               + \frac{\dot{h}}{2}\right\} \nonumber \\ &-&   3H(\clam-w)\delta,
\eea
\be{thetaevol}
\dot{\theta} = -H(1-3\clam)\theta+\frac{\clam k^2}{1+w}\delta-k^2\sigma,
\ee
where $h$ is the trace of the synchronous metric perturbation.
Here $\sigma$ is the anisotropic stress of dark energy, related to notation
of Eq.(\ref{fluid}) by $(\rho + p)\sigma \equiv
-(\hat{k}_i\hat{k}_j-\frac{1}{3}\delta_{ij})\Sigma^{ij}$. From the above equations it is then clear that, while
$w$ and $\clam$ determine respectively the background and perturbative
pressure of the fluid that is rotationally invariant, $\sigma$ quantifies
how much the pressure of the fluid varies with direction. 

To close the system of equations,  we describe the
evolution of the anisotropic stress with an equation adopted from Hu \citep{Hu:1998kj}, 
\be{sigmaevol}
\dot{\sigma}+3H\sigma = \frac{8}{3}\frac{\cvis}{1+w}(\theta+\frac{\dot{h}}{2}+3\dot{\eta}).
\ee
This parameterization leads to reasonable results and approximates the evolution of any
fluid present in the standard cosmological model, in particular neutrinos and photons which
have a non-zero anisotropic stress \citep{Hu:1998tj}. More specifically, for those relativistic components, the correct choice for
the viscous parameter is $\cvis=1/3$. A perfect fluid (vanishing shear viscosity)  should have $\cvis=0$.

Equivalently, one can describe the physical properties of $\sigma$ by introducing the rescaled parameter $\avis = \cvis/(1+w)$. While
$\cvis$ is somehow physically analogous to sound speed squared, the $\avis$ is the quantity directly multiplying the
source term of the stress (see RHS of eq. (\ref{sigmaevol})). In this
study we will use both the $\cvis$ and $\avis$ parameterizations. Since $w$ is constrained near $w=-1$, where the
relation between these parameters is divergent, using one or the other might lead to different results and interpretations. The statistical
details also depend on which parameter one assumes a uniform distribution, and it is useful to test how robust ones conclusions
are to such assumptions. 

Usually, a dark energy fluid with nonzero $\avis$ generates shear stress which tends to smoothen its distribution. However, the consequences to phantom dark energy
are qualitatively different and for such a fluid, with $w<-1$, a shear stress drives the clustering. With negative $\avis$,
exponential growth is typical for all kinds of dark energy. In the case of positive $\avis$, and constant $w$ and $\clam$, 
the effects are confined to superhorizon scales and typically small \citep{koivisto:2005}. 
This is in contrast to anisotropic stress which originates from modifications of the gravity sector or quintessence couplings, since they typically 
modify the gravitational potentials at small scales. A recent study by Amendola {\it et al} \citep{Amendola:2007rr} considers the possibility of using  weak lensing to obtain 
limits on the dark energy parameters motivated partially by modified gravity \citep{Amendola:2007rr}. Our study can be considered as an exploration of complementary 
aspects since the effects we encounter, occur (except for negative $\clam$ or $\avis$) at some orders of magnitude larger scales than 
those possible to probe with weak lensing experiments. An interesting approach is also that of Caldwell {\it et al} \citep{Caldwell:2007cw} who,  
parameterizing directly the deviation from the general relativistic perfect fluid metric, and assuming it to depend on the amount of dark 
energy, find that this assumption leads to relatively weak constraints on cosmological scales, whereas the bounds on the solar system are known 
to be tight.

\begin{figure}
\center{
\includegraphics[width=9cm]{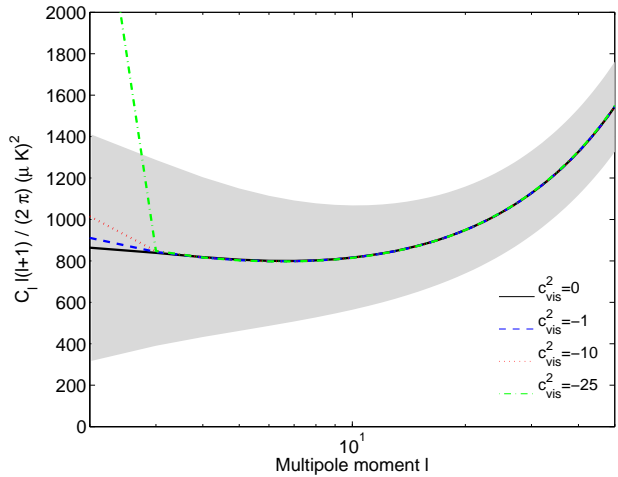} 
\includegraphics[width=9cm]{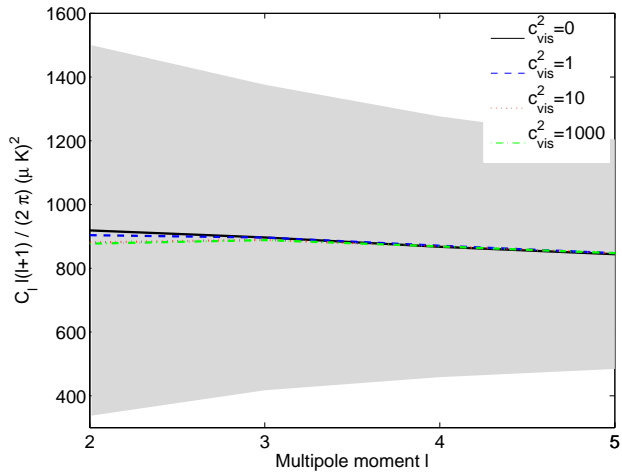}
\caption{\label{fig:cls_cvis} The CMB temperature power spectrum for
  models with different values of $w$ and $\cvis$. The upper panel is
  for a model with $w=-1.2$ and $\cvis \leq 0$, while the lower panel shows the results
  for a model with $w=-0.8$ and $\cvis \geq 0$. The grey shading
  indicates the cosmic variance
  around the models with $\cvis = 0$. All models plotted here have
  $\clam=1$.}}
\end{figure}

To illustrate the effect of $\cvis$ on the CMB power spectrum, we have
in Figure \ref{fig:cls_cvis} shown the CMB temperature power
spectra for models with different values of $w$ and $\cvis$. In models with
  $w=-1.2$, one sees that the deviation from the $\cvis=0$ model starts to become significant with
$\cvis = -25$. With $w=-0.8$, the deviations remain small, even with
$\cvis=1000$. From this we would expect that it is possible to find
some lower limit on $\cvis$ in models with $w<-1$, while  it will be
difficult to find upper limits on $\cvis$ in models with $w>-1$.
\begin{figure}
\center
\includegraphics[width=9cm]{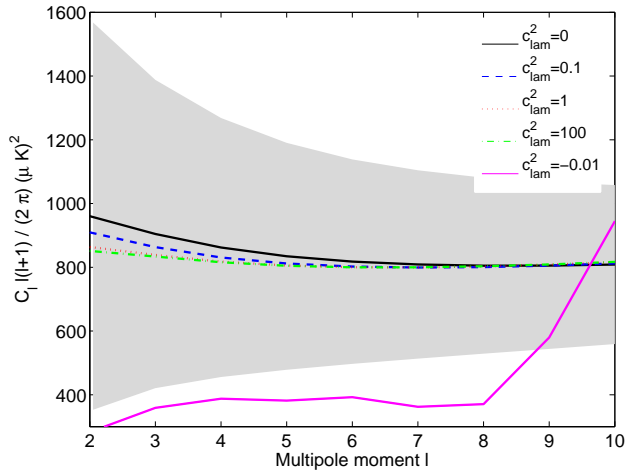}
\includegraphics[width=9cm]{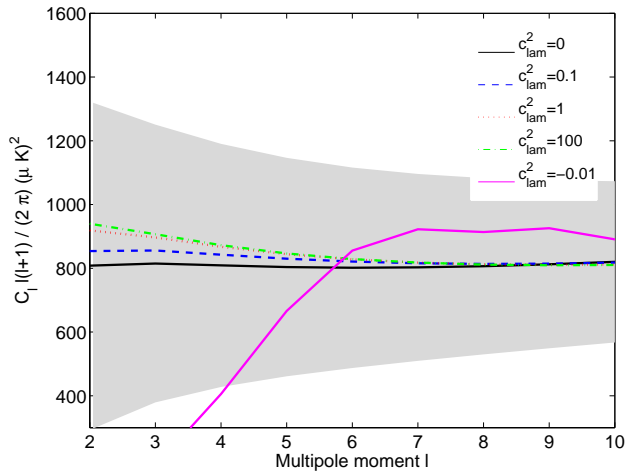}
\caption{\label{fig:cls_clam} The CMB temperature power spectrum for
  models with different values of $w$ and $\clam$. The upper panel
  shows the results for a model with $w=-1.2$, while the lower panel
  shows models with $w=-0.8$.  The grey shading
  indicates the cosmic variance. All  models shown here $\cvis=0$.}
\end{figure}

In Figure \ref{fig:cls_clam} we have plotted the CMB temperature power
spectra for models with different values of $w$ and $\clam$. We see
that for values of $\clam$ increasing over $\clam=1$, the shape of the
power spectra seems to stabilize, even when $\clam$ gets as large as
$\clam=100$. So also in the case of $\clam>0$ it looks difficult to distinguish
between the different models. 
Such difficulty can be understood noticing that, increasing $\clam$ will lead to a smoothening of a quintessence-like case. This is itself 
already almost smooth. Therefore we do not see much difference even when $\clam=100$. Notice, however, that negative $\clam$ is also conceivable, and
it corresponds to gravitational instability. This leads to the
clustering of dark energy and thus to possibly observable effects. In
Figure \ref{fig:cls_clam} we see that the observable effect of $\clam<0$ is much
more evident than for $\clam>0$. Such possibility 
will also be investigated in the following sections.

\section{Data and methods}
\label{data}

In our analysis we use data from observations of  the anisotropies of the Cosmic Microwave Background, the distribution of large
scale structures (LSS), the  distance-redshift relation from type Ia supernovae (SNIa), and an additional prior on the Hubble parameter. The parameter 
estimation analysis is done using a modified version of the Markov chain Monte Carlo code CosmoMC \citep{lewis:2002}. 

In our basic cosmological model, we assume a flat, homogeneous and isotropic
background spacetime, and a simple power law primordial
power spectrum. Thus, we allow the six parameters $\{ \Omega_bh^2,
\Omega_m, \log(10^{10}A_S), h, n_s, \tau  \}$ to vary freely. Here
$\Omega_i$ is the ratio of the energy component $i$ to the critical
density today, $A_S$ is the amplitude of the primordial power
spectrum, $h$ is the dimensionless Hubble parameter today, $n_s$ is the
scalar spectral index and $\tau$ is the optical depth at
recombination. The exact
definitions of the parameters are given by the CosmoMC code. In
addition to these basic six parameters we have also studied changes in
$w$, $\cvis$ (or $\avis$) and $\clam$.

The most constraining CMB dataset at present is the 3-year data
release from the WMAP team. In our analysis we have used both the
temperature \citep{hinshaw:2006} and polarization \citep{page:2006} data
from this experiment together with the likelihood code provided by the
WMAP team \footnote{http://lambda.gsfc.nasa.gov; version v2p2.}. Since
the effects we are studying here are most prominent on large angular
scales in the CMB signal, we do not take additional small-scale CMB
experiments into account in this analysis.   

In one case we have also utilized LSS data from the Sloan Digital Sky
Survey Luminous Red Galaxy Sample (SDSS-LRG) \citep{tegmark:2006}. When
using information on SNIa distance-redshift relation, we use data from
the Supernova Legacy Survey (SNLS) \citep{astier:2006}. 

We have also used a prior of $h=0.72 \pm 0.08$ from the Hubble
Space Telescope Key Project (HST) \citep{freedman:2001} and a top-hat
prior on the age of the universe, 10Gyr$<$Age$<$20Gyr, throughout the
entire analysis.   


\section{Constraints from present data}
\label{cons1}

\subsection{Constraining dark energy viscosity}
\label{cons_vis}

To start with we have used the $\cvis$ parameterization, looking at two
distinct scenarios. In one case we have  $\{ w>-1, \cvis>0 \}$, and in
the other case  $\{ w<-1, \cvis<0 \}$. Here, and for the rest of
subsection \ref{cons_vis}, we have set $\clam=1$. 

\begin{figure}
\center
\includegraphics[width=9cm]{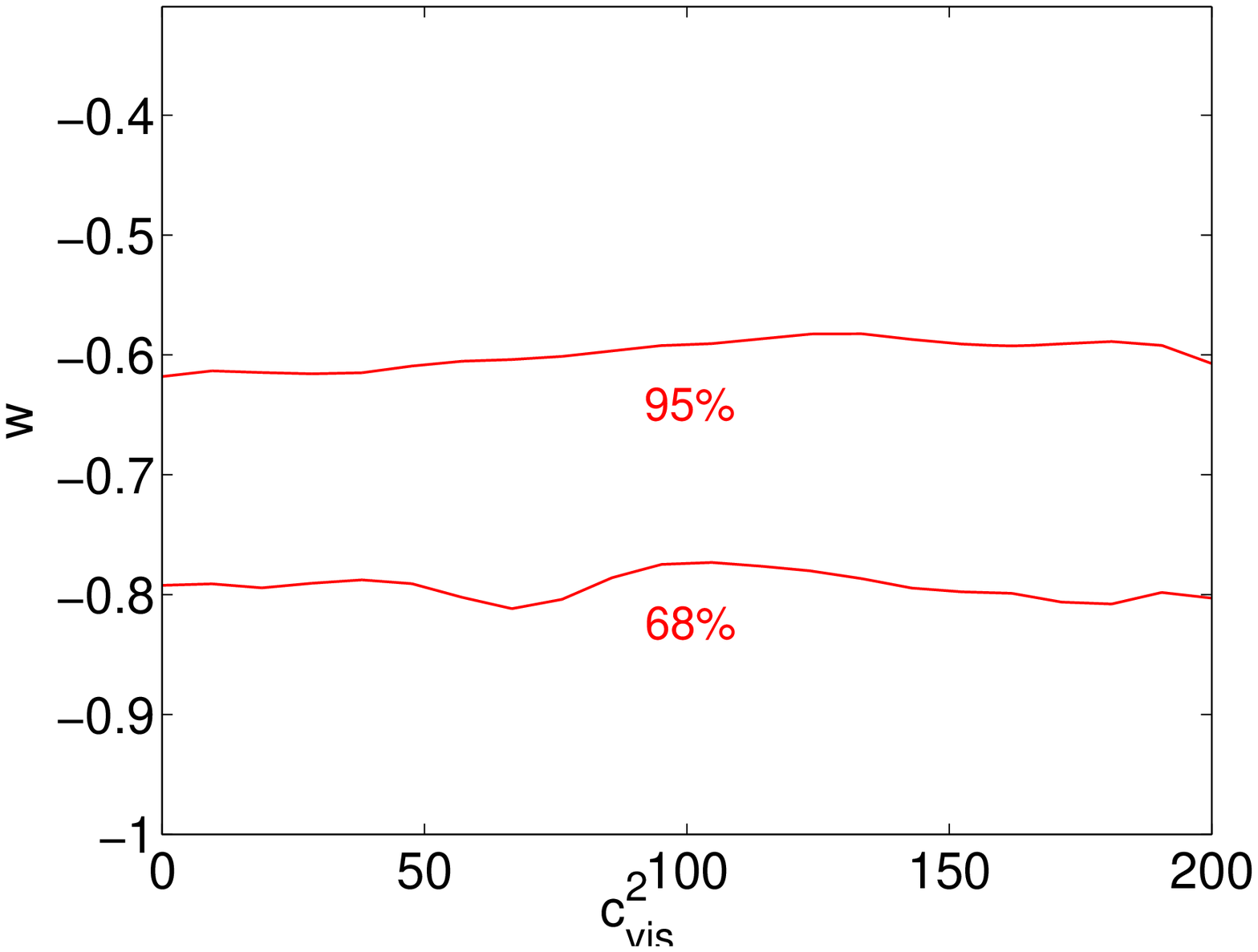} 
\includegraphics[width=9cm]{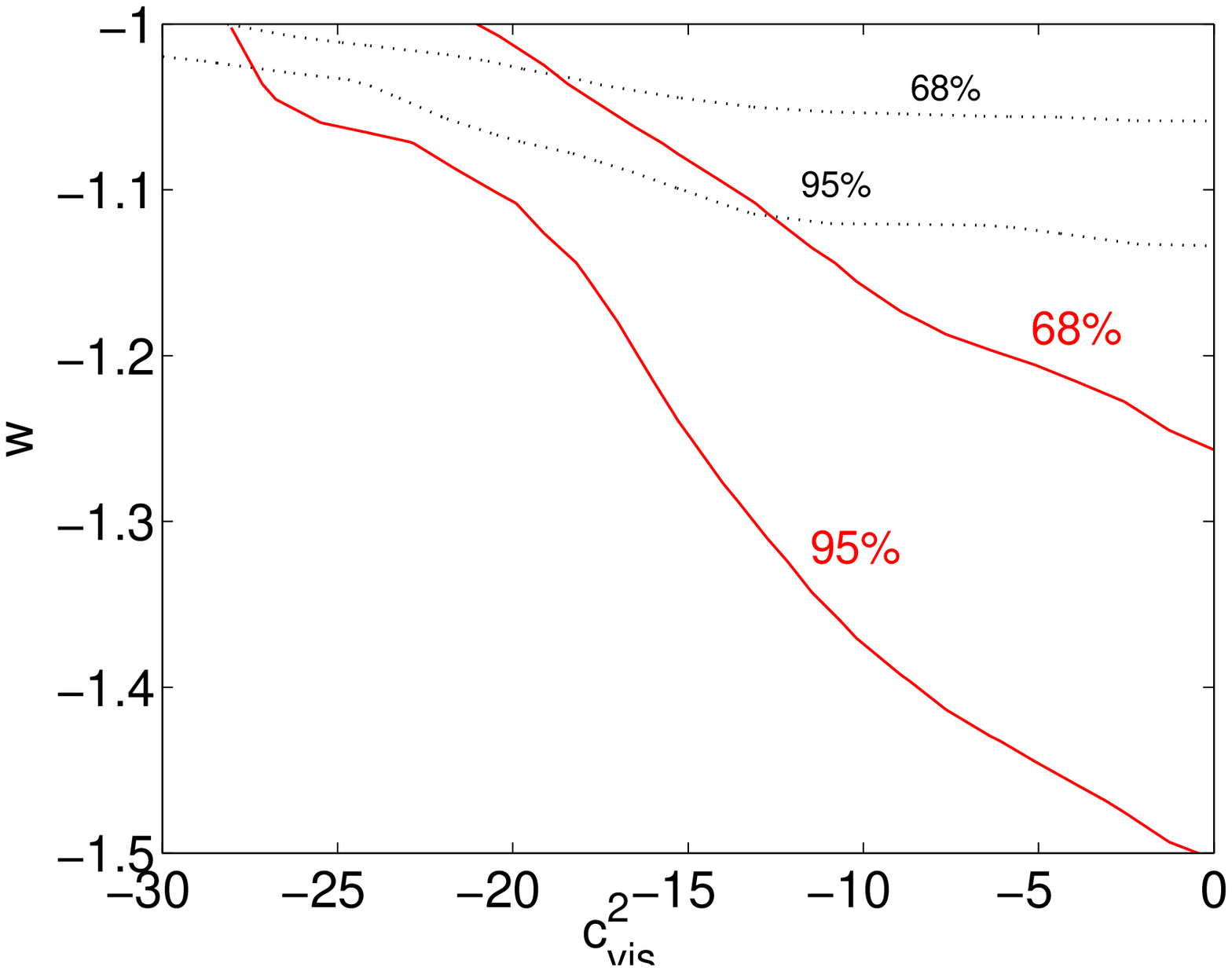}
\caption{\label{fig:cvis_w} 68\% and 95\% contours of $ \cvis $ and
  $ w $. In the upper panel $ w>-1 $ and $ \cvis>0 $, while in the lower panel
  $ w<-1 $ and $ \cvis<0 $. We see that the constraints are much tighter
  for negative $ \cvis $ than for positive. In the right plot we have
  also shown how the constraints improve when including SDSS-LRG and
  SNLS data (dotted black contours). In these models $\clam=1$.}
\end{figure}

In Figure \ref{fig:cvis_w} we show the 65\% and 98\% confidence level (CL)
contours in the $w-\cvis$ plane for the two main scenarios mentioned
above. We see that in
the first case there is no clear degeneracy between $\cvis$ and $w$,
and we find no upper limit on $\cvis$ here. This is exactly what we
would expect from an inspection of Figure \ref{fig:cls_cvis}; when
$\{ w>-1, \cvis>0 \}$, the effects of varying $\cvis$ are negligible. 

For the latter case, where
$w<-1$ and $\cvis<0$, we do however find a degeneracy between these
parameters, and in this case we also find lower limits on
$\cvis$. Using only WMAP data we find $\cvis>-19.5$ at 95\% confidence
level (CL) from the 1 dimensional distribution of $\cvis$. 
Since there are observable effects of a non-zero $\cvis$
in this scenario, we also tried to add SDSS-LRG and SNLS data to see
if this would improve our constraints on $\cvis$. In this case the lower limit on $\cvis$
becomes $\cvis>-24.9$ at 95\% CL. We can also easily see from the figure that adding SDSS-LRG and SNLS data does
not improve the limits on $\cvis$, only on $w$. The reason is that the effects of $\cvis$  occur
at larger scales than probed by the SDSS-LRG. Therefore this
additional data would only be interesting if it could break
degeneracies between $\cvis$ and other parameters that govern the
shape of the power spectrum for low $l$s. We see from Figure
\ref{fig:cls_cvis} that although we do have a degeneracy between
$\cvis$ and $w$, the limits on $\cvis$ would only be improved by some
data set that would favor a $w<-1$.

\begin{figure}
\center
\includegraphics[width=9cm]{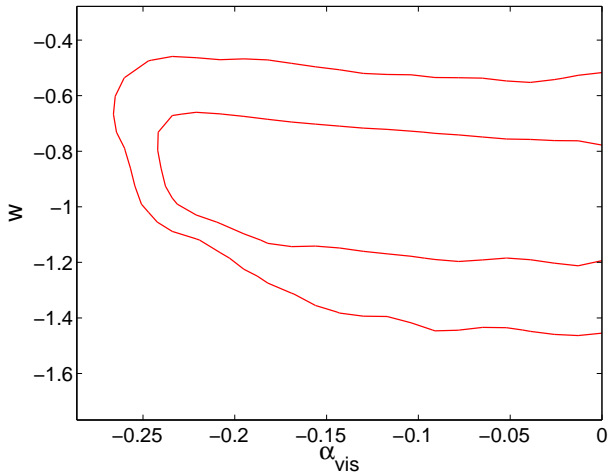}
\caption{\label{fig:alpha_w} 68\% and 95\% contours of $\avis$ and
  $w$. We see that slightly negative values of $\avis$ are allowed,
  and that the allowed values for $\avis$ in this parameter range has
  only a weak dependence on $w$. In this model $\clam=1$.}
\end{figure}

Next, we turn to the $\avis$ parameterization. To constrain this
parameter we focus on a model where $\avis <0$, since we find no
upper limits on a positive $\avis$. 
In Figure \ref{fig:alpha_w} we show the 65\% and 98\% CL
contours for $w$ and $\avis$ in this model. We see that $\avis$
is bounded from below and that values of $\avis$ slightly below zero
are allowed. In this case we find that $\avis> -0.23$ at 95\%
CL.

\begin{figure}
\center
\includegraphics[width=9cm]{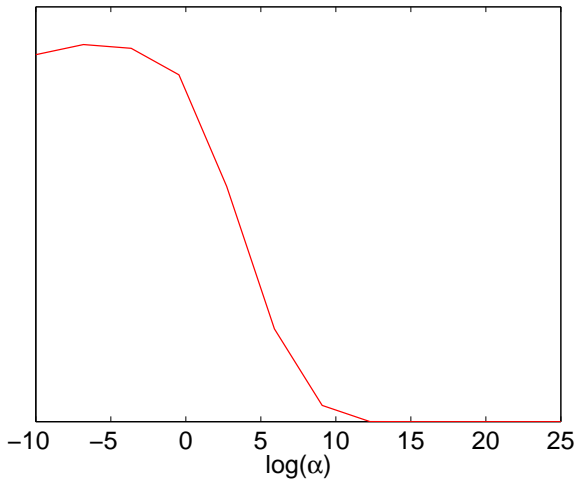}
\caption{\label{fig:alpha_1D} Marginalized probability distribution for
  $\log(\avis)$ for a model with $w=-1.2$ and $\clam=1$.}
\end{figure}

Since we got a lower limit on $\cvis$ when having $w<-1$, we would
also expect to find an upper limit on $\avis$ in this case. In Figure \ref{fig:alpha_1D} we show the 1 dimensional marginalized
likelihood for $\log \avis$ for a model with $w=-1.2$. We find that
$\log \avis < 3.2$ at 95\% CL ($\avis<24.5$), which corresponds well
to the limits we found for $\cvis$ with $w<-1$.  

\subsection{Constraining dark energy sound speed}

In this section we investigate what constraints can be put in the dark energy
sound speed, $\clam$, from WMAP data. Throughout this subsection $\avis=0$.

From Figure \ref{fig:cls_clam} we see that the effect of varying
$\clam$ away from $\clam=1$ is negligible for models with $\clam>1$,
but notable when varying $\clam$ in the regime $\clam<1$. Thus we
would expect to find a lower limit, but not an upper limit on the
value of $\clam$, and this is indeed the case.

\begin{figure}
\center
\includegraphics[width=9cm]{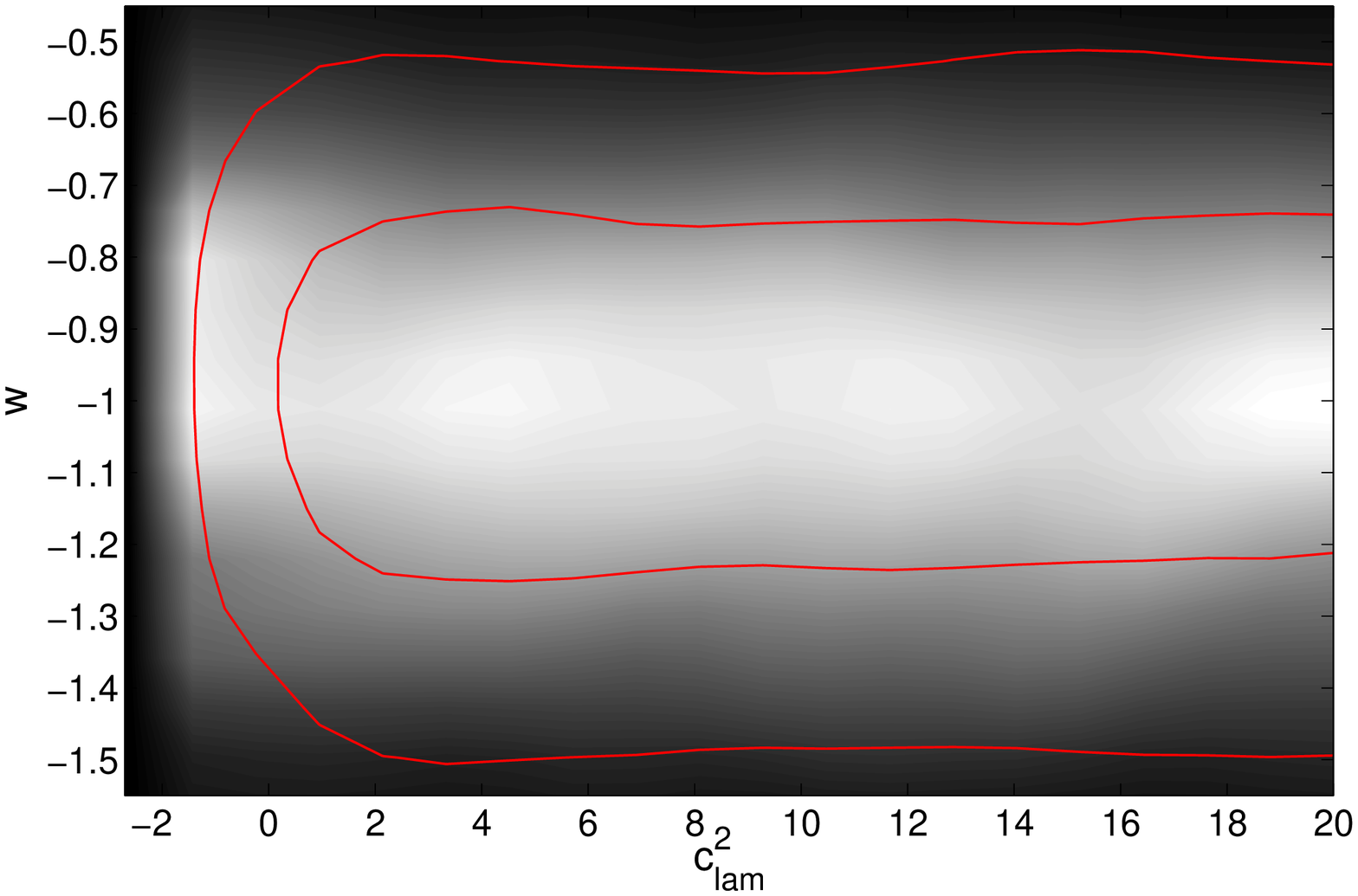}
\includegraphics[width=9cm,height=5.9cm]{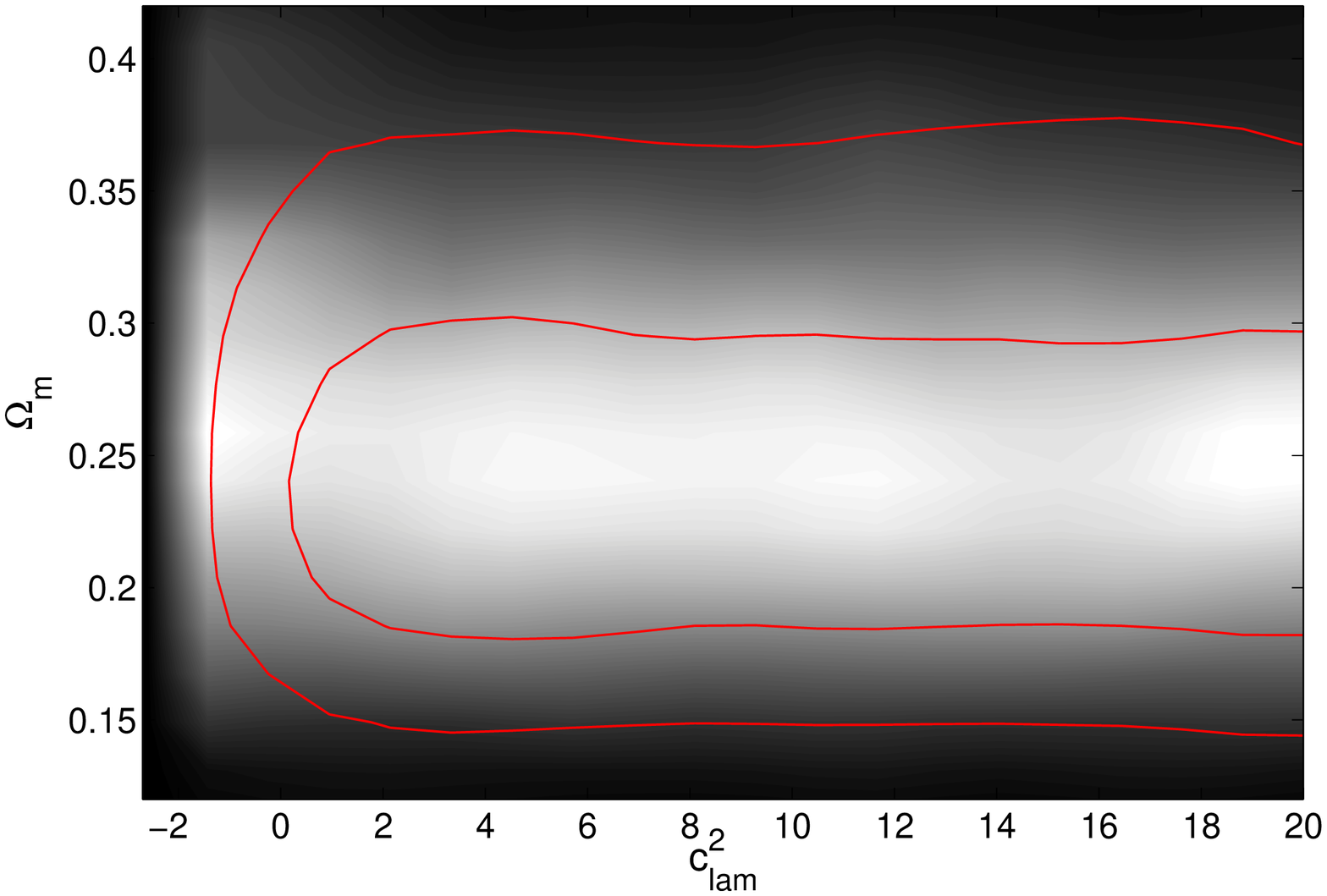}
\caption{\label{fig:clam_dobbel} 68\% and 95\% confidence contours in
  the $\clam$-$w$ and $\clam$-$\Omega_m$ planes using WMAP data. We
  see that the degeneracies between $\clam$ and both these parameters
  are very weak. The colored background indicates the mean likelihoods
  of the samples. In this model $\avis=0$.}
\end{figure}

In Figure \ref{fig:clam_dobbel} we have plotted the 2-dimensional
confidence contours in the $\clam-w$ and $\clam-\Omega_m$ planes that
we get from the WMAP data and for a model where $\clam$ is allowed to
vary freely in the interval $\{-20,20\}$. In both cases we see that we do get a lower
limit, and also that there is hardly any degeneracy  between
$\clam$ and these parameters (or any of the other parameters in our model). Therefore, as in the case of $\cvis$ and
$\avis$, we would not expect to improve our constraints on $\clam$ by
using other types of cosmological data. 

Note that the intuitively plausible models with $\clam=0$ reside
at the lower end of a long ``ridge'' of $\clam$ values that fit the
data almost equally well. Therefore the interpretation of the confidence
contours should be done with some care. This is the reason why we do
not state any lower limit on $\clam$ here. Also note that the shaded
backgrounds in Figure \ref{fig:clam_dobbel} refer to the \emph{mean
  likelihood} of the samples, whereas the contours show the
\emph{marginalized likelihoods}. For such highly non-gaussian distribution
that we have here, there is a significant mismatch between these. See
Appendix C in \citep{lewis:2002} for a discussion on this.

\begin{figure}
\center
\includegraphics[width=9cm]{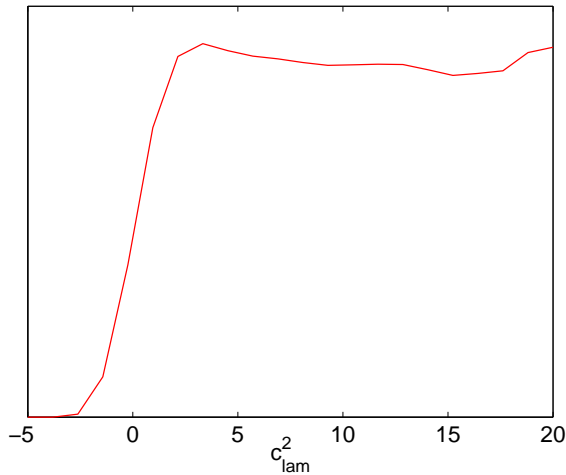}
\caption{\label{fig:clam_1D} The marginalized probability
  distribution for $\clam$ using WMAP data. No upper limit is
  obtained. In this model $\avis=0$ and $w$ is allowed to vary freely.}
\end{figure}

In Figure \ref{fig:clam_1D} we show the marginalized probability
distribution of $\clam$ for the same model. In the figure we have also
indicated the mean likelihoods of the samples by the background
shading color. Here we clearly see the
effect of the converging power spectrum for $\clam>1$, as the
likelihood for models with $\clam>1$ does not decrease. We also see
that we are allowed to have models with $\clam$ slightly negative, but
that the models rapidly become incompatible with the data for $\clam<0$.

\subsection{Effect on other parameters}

Another interesting issue is whether the extra freedom in the dark
energy fluid will affect the constraints on the other parameters in our
cosmological model. That is, are the parameter constraints in
the $\Lambda$CDM model robust to changes in $\cvis$ and $\clam$. This was also studied in \citep{ichiki:2007}, where they found
that $\avis$ did not change the constraints in the other cosmological parameters
significantly, but that varying $\clam$ shifted the other parameter
ranges slightly. 

\begin{figure}
\begin{center}
\includegraphics[width=9cm]{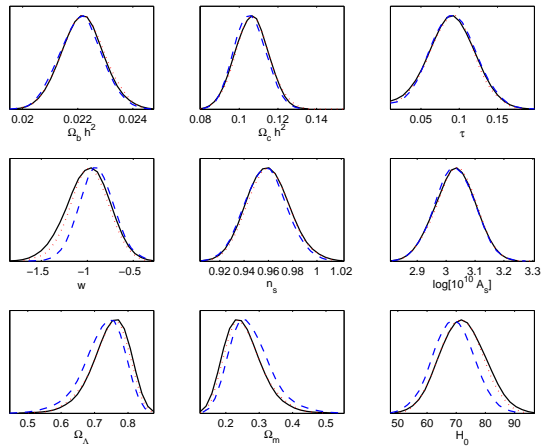}
\caption{\label{fig:1D} Marginalized parameter distributions in
  three models with free $w$. The solid black lines correspond
  to a model with $\avis=0$ and $\clam=1$. The dotted red lines show the results
  from a model where $\avis$ is allowed to vary freely between $-20$ and
$20$ and $\clam =1$. The dashed blue lines show the constraints from a
model with $\alpha_{vis}=0$ and $c_{lam}=0$.}
\end{center}
\end{figure}

In Figure \ref{fig:1D} we have plotted the marginalized
likelihoods for different cosmological parameters in the case of a 7
parameter model with free $w$ but with $\avis = 0$ and $\clam=1$. We
compare this to a model where $\avis$ is allowed to vary freely
in the interval $\{-20,20\}$ (the same model as shown
in Figure \ref{fig:alpha_w}). Also shown is a model with $\avis=0$ and
$\clam = 0$. We see that the extra freedom in the
$\avis$ parameter does not change the other parameter distributions
significantly. We do however get a slight shift in the parameter
distributions by changing $\clam$ from 1 to 0. This is consistent with
the results from \citep{ichiki:2007}. 

The most notable effect of changing from a model with $\clam=1$ to a
model with $\clam=0$, is that the probability distribution for $w$
becomes narrower in the latter case. For a model with $\clam=1$ we
find $w=\{-1.47,-0.57\}$ at 95\% CL, while this range changes to
$w=\{-1.26,-0.52\}$ for a model with $\clam=0$. For the other
parameters, the effect of changing $\clam$ is not very notable.

\section{Constraints from future data}
\label{cons2}

As we have seen, only weak constraints can be found on the $\cvis$,
$\avis$ and $\clam$ parameters using present data. We have also argued
that the effect of including other types of data sets, like LSS and
SNIa, would not be very helpful, as such kinds of data sets are not
affected significantly by these parameters. Also, unless $w$ is
significantly below -1, they will not serve to break parameter
degeneracies for the parameters studied here. Will it then be possible
to improve our constraints with future CMB data?

To answer this question we have simulated a ``perfect'' CMB temperature data set, where the
error bars
are defined only from cosmic variance (CV) around a power spectrum generated
from the best-fit $\Lambda$CDM model (with $w=-1$, $\cvis=0$ and $\clam=1$) from WMAP data. The likelihood part of CosmoMC
has been modified to
use this perfect data instead of the WMAP measurements. The likelihood is calculated
as in \citep{verde:2003}:
\begin{equation}
 -2 \log \mathcal L(x_i; p_i) = \sum_\ell (2\ell +1)
\left[
\ln \left(\frac{C_\ell^{model}}{C_\ell^{sim}}\right) + \frac{C_
\ell^{sim}}{C_\ell^{model}} -1
\right]
\end{equation}
For this mock data set we have used multipoles from $l=2$ to $l=2000$
in our analysis, and also here we have added the same prior on $H_0$
and age as earlier. A completely noise-free data set is of course not
realistic. However, it is an interesting case to study, since effects
that cannot be seen here, will never be possible to see using a real
CMB temperature experiment. Note that, when using this mock data set,
we do not include any polarization data in our analysis. 

\begin{figure}
\center
\includegraphics[width=9cm]{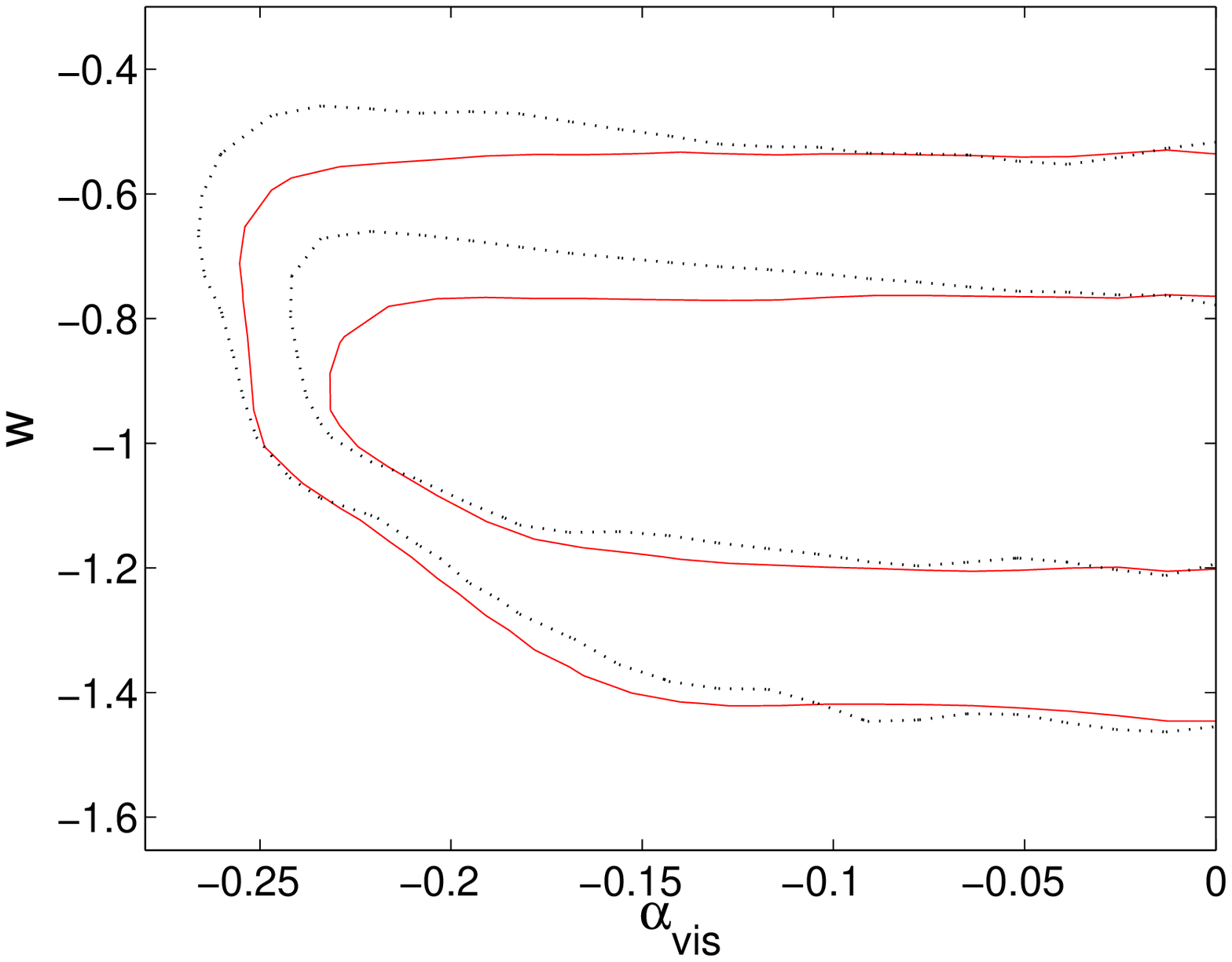}
\caption{\label{fig:alpha_w_p} 68\% and 95\% contours of $\avis$ and
  $w$ in a model with $\clam=1$ and using only "perfect" CV limited CMB
  temperature  data (solid lines). The dotted lines shows the results
  obtained when using WMAP3 data, as shown in Figure
  \ref{fig:alpha_w}.}
\end{figure}

In Figure \ref{fig:alpha_w_p} we show the constraints in
the $\avis$-$w$ plane in a model with $\clam=1$ and $\avis<0$. This is compared with
the results from using WMAP data alone, as also shown in Figure
\ref{fig:alpha_w}. As we can see, even in this idealized case, we do
not see any major improvement in our constraints on $\avis$. In this
case the lower limit increases from $\avis>-0.23$ (from WMAP data) to
$\avis>-0.22$. 

\begin{figure}
\center
\includegraphics[width=9cm]{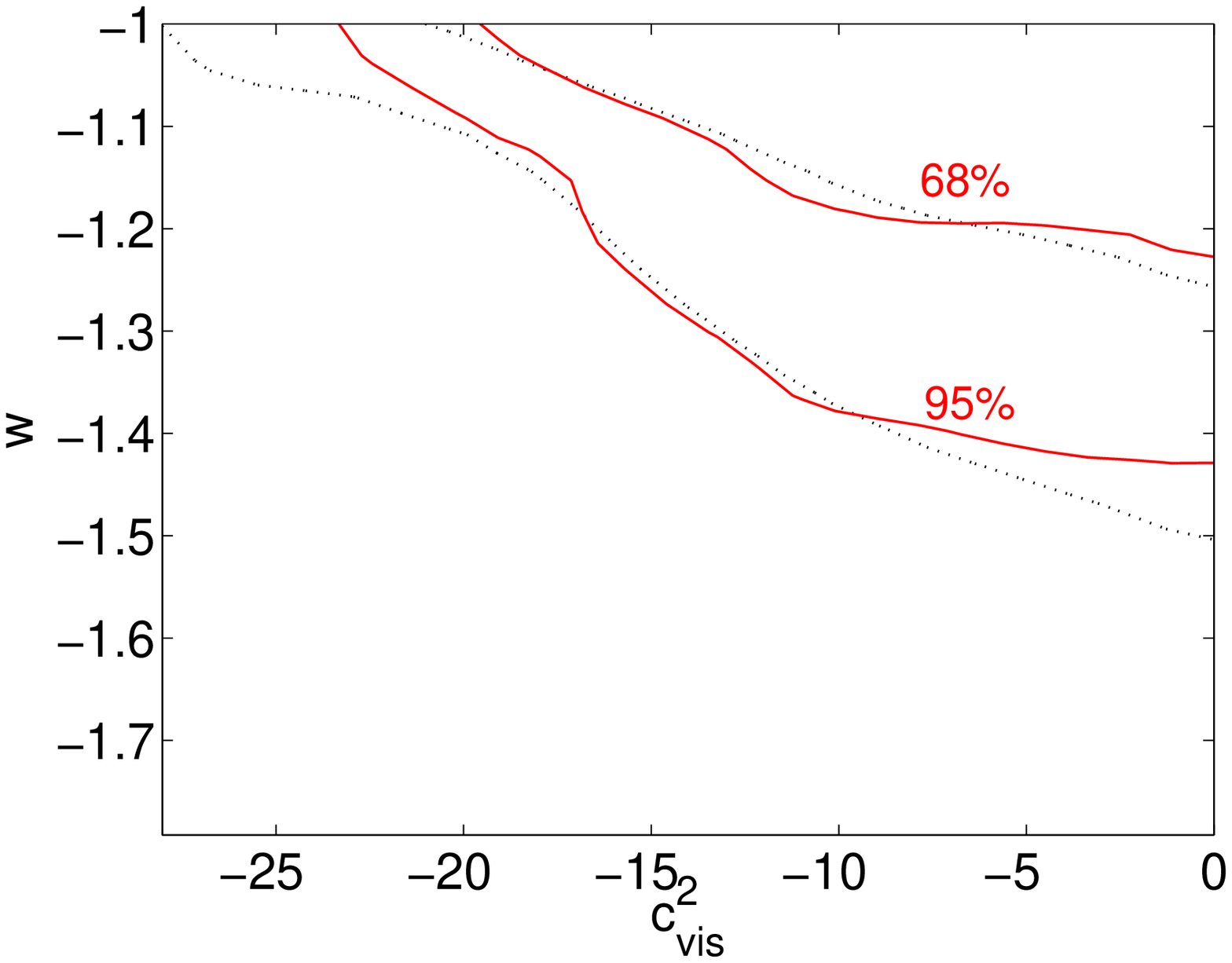}
\caption{\label{fig:cvis_w_p} 68\% and 95\% contours of $\cvis$ and
  $w$ in a model with $\clam=1$ and using only CV limited CMB
  temperature data (solid red lines). The dotted lines shows the results
  obtained when using WMAP3 data, as shown in Figure
  \ref{fig:cvis_w}.}
\end{figure}

In Figure \ref{fig:cvis_w_p} we have used the CV limited mock data to redo one the most interesting
 case from the analysis with WMAP data, namely  the constraints
in the $\cvis$-$w$ plane with $w<-1$ and $\cvis<0$, as shown in Figure
\ref{fig:cvis_w}. We see that the constraints in this area improve
slightly, but not very significantly. Using the CV-limited mock data
we find $\cvis>-17.5$, compared to $\cvis>19.5$ using WMAP data.

\begin{figure}
\center
\includegraphics[width=9cm]{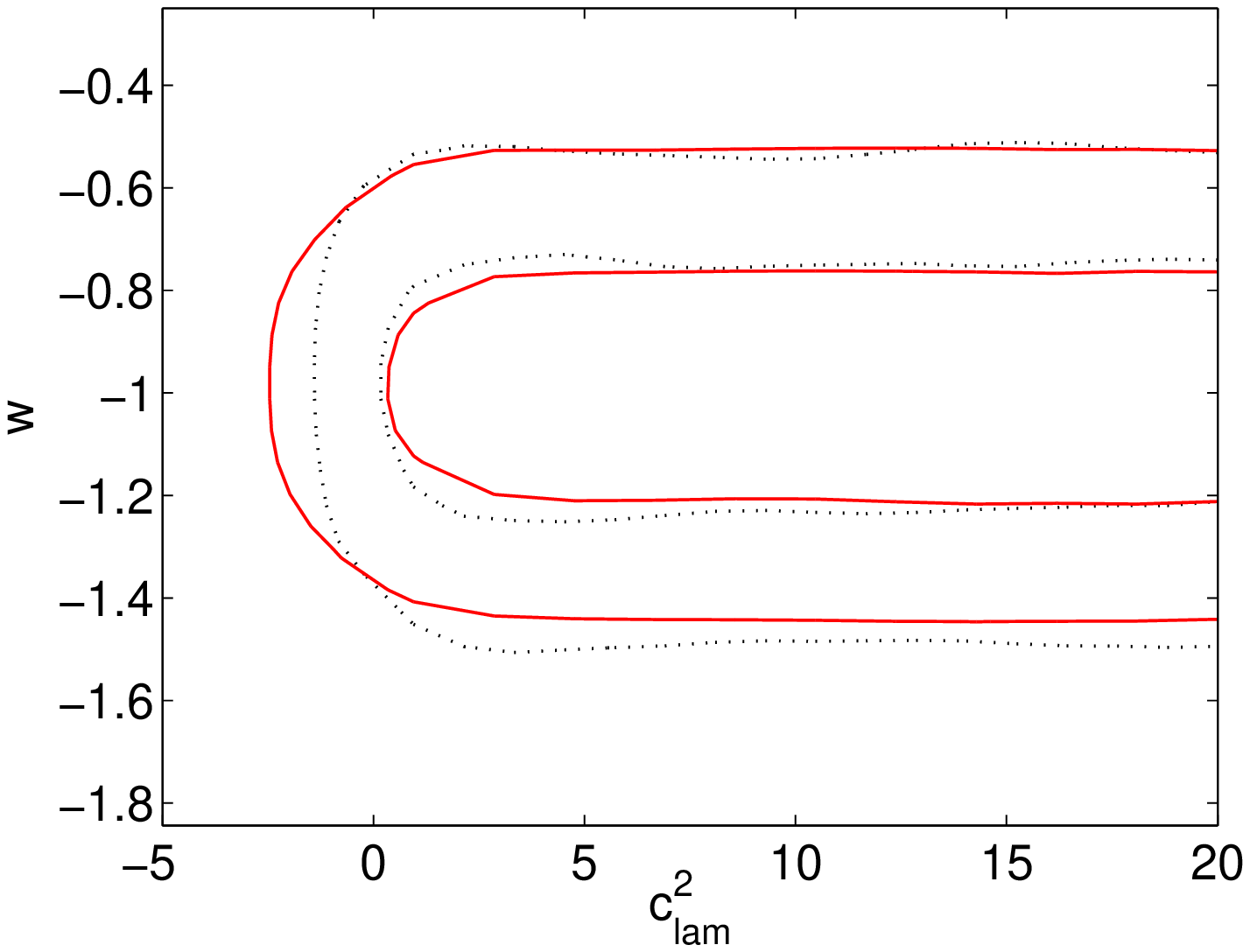}
\includegraphics[width=9cm]{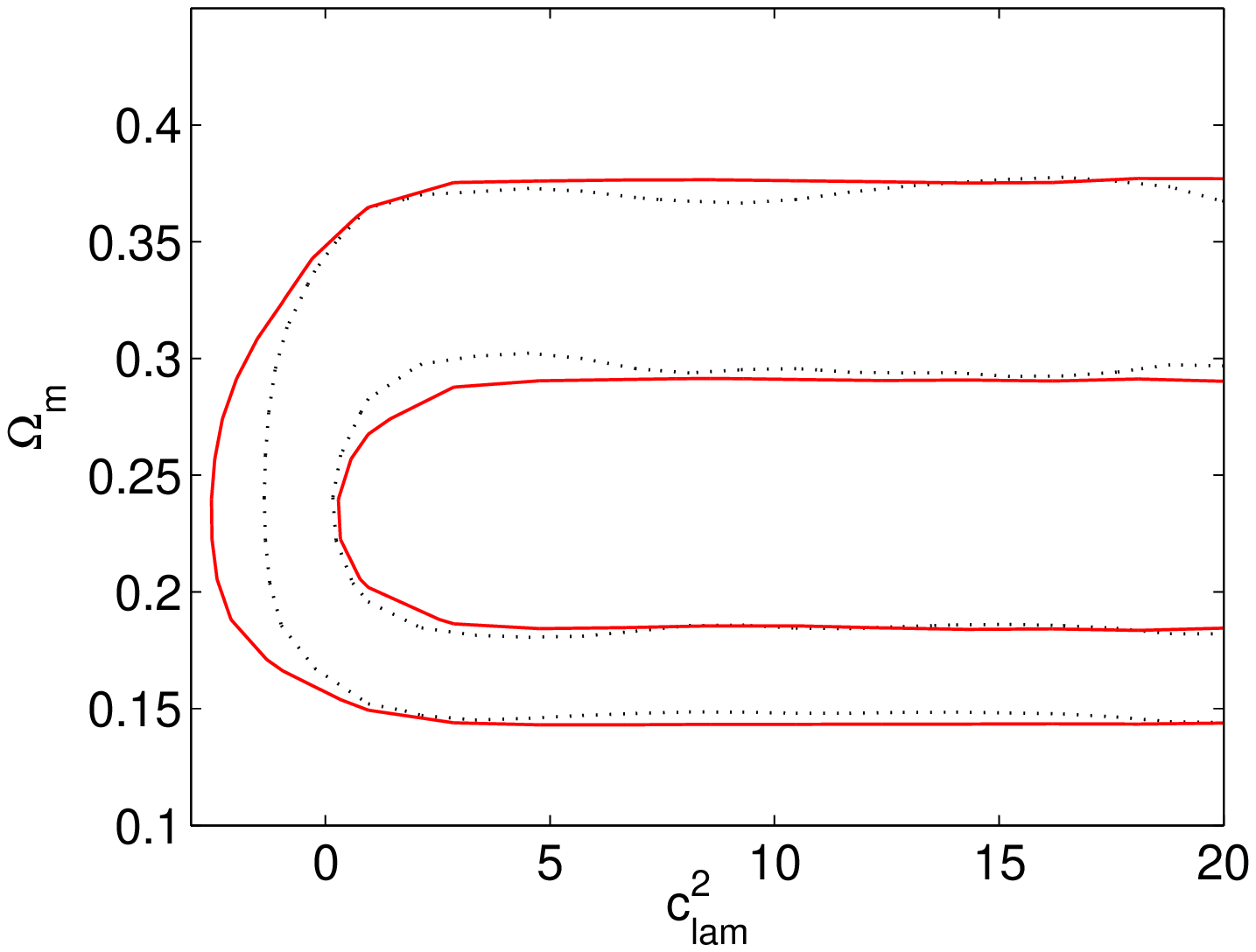}
\caption{\label{fig:clam_dobbel_p} 68\% and 95\% confidence contours in
  the $\clam$-$w$ and $\clam$-$\Omega_m$ planes using only CV limited
  CMB temperature data (solid red lines). The black, dotted contours are the
  results obtained from using WMAP data, as shown in Figure
  \ref{fig:clam_dobbel}. The constraints are not improved when using
  ``perfect'' data. In this model $\avis=0$.}
\end{figure}

\begin{figure}
\center
\includegraphics[width=9cm]{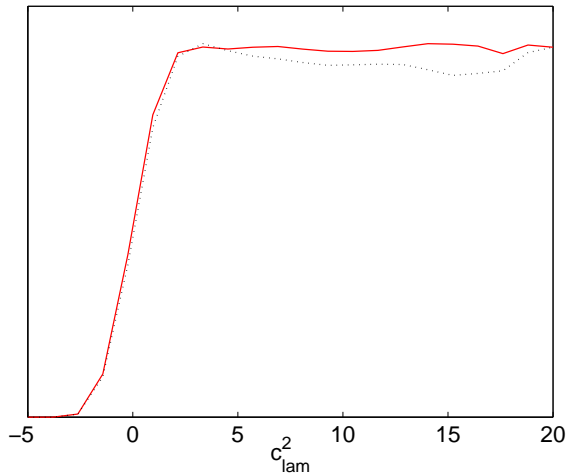}
\caption{\label{fig:clam_1D_p} The marginalized probability
  distribution for $\clam$ using only CV limited CMB temperature data
  (solid red line). The black, dotted line indicates the result from
  using WMAP data. The constraints are not improved by using
  ``perfect'' data. In this model $\avis=0$ and $w$ is allowed to vary
freely.}
\end{figure}

We have also redone the constraints on the dark energy sound speed,
$\clam$, using the ``perfect'' data set. In Figure
\ref{fig:clam_dobbel_p} we show the confidence contours in the
$\clam$-$w$ and $\clam$-$\Omega_m$ planes, as done earlier with WMAP
data. In Figure \ref{fig:clam_1D_p} we have the corresponding
1-dimensional probability distribution. As we can see, using the
perfect CMB temperature data, does not improve our constraints on this
parameter at all.  

That improved CMB data does not help to constrain dark matter
properties, is not very surprising. As seen in Figures
\ref{fig:cls_cvis} and \ref{fig:cls_clam}, effects occur mainly on
large scales, where the error bars in the WMAP data already are
dominated by cosmic variance.  

\section{Discussion and conclusions}
\label{conc}

We have investigated  the possibility of using cosmological observations to constrain a possible dark energy anisotropic stress.
We have parameterized the dark energy component in a very general way. Such generalized cosmological fluid is characterized by 
an equation of state ($w$), a speed of sound ($\clam$) and an anisotropic stress ($\avis$ and $\cvis$). We have subjected this 
parameterization to the most detailed and most extensive scrutiny this far. 

We have found that it is difficult to constrain the properties of an
imperfect dark energy fluid. If the Universe resides in the phantom
regime, with $w<-1$, one can put a weak lower limit on
$\cvis$. In this case we found $\cvis>-19.5$ at 95\% CL. 
However, $\cvis$ is not bounded from above. We also found that in a model with negative
$\avis$, this parameter can be constrained to $\avis>-0.23$ at 95\%
CL using WMAP data. We found no upper limit on $\avis$ in models with $\avis>0$.
These results reflect the fact that a dark energy fluid with nonzero $\avis$ 
generates shear stress which tends to smoothen its distribution. However, the 
consequences to phantom dark energy are qualitatively different and for such 
a fluid, with $w<-1$, a shear stress drives the clustering.

A similar conclusion applies to the dark energy sound speed, $\clam$, where one 
also can put some lower limits, but no upper limits on its value.  

Considering a simulated ``perfect'' CMB data set,
only limited by cosmic variance, we found that improved CMB data sets
are not likely to improve these constraints any further in the future.

These results stem from the fact that there is always an area in the parameter space describing the possible properties of dark energy, where the 
fluid is different from the cosmological constant or quintessence, but observationally indistinguishable. Optimistically though, our Universe 
could happen to reside in some other area of this parameter space. Indeed, in some special cases, one might be able to find a detect a possible 
imperfectness of the dark energy fluid. 

One could consider dark matter and dark energy into a single imperfect fluid \citep{koivisto:2005}. Since then the effects of shear 
viscosity are present already in the earlier evolution of universe, they in general have observable consequences. This is somewhat related to models
featuring the presence of early dark energy \citep{Doran:2006kp} (which is excluded by our single-parameter description of the evolution of the dark 
energy equation of state). It could be therefore interesting to study how the previous constraints derived on the presence of early dark energy  will 
be modified when one takes into account possible imperfect properties of the fluid.

The anisotropic stress of dark energy can have clear signatures in the case that it does not average exactly to zero at large scales. 
Then the nonzero $\sigma$ does not only appear as a statistically isotropic perturbation of the fluid, but will drive the overall expansion of the 
universe anisotropically. Clearly the resulting modifications of the Friedmann equation can be tightly constrained \citep{Koivisto:2007bp}.

However, such specific cases aside, the arguably simplest dark energy extension of the $\Lambda$CDM model introduces an equation of state $w \neq 1$.
Inevitably, the more general model then generates dark energy perturbations, thus forcing one to specify also the two parameters $\clam$ and $\cvis$. 
Again, arguably the simplest parameterization then considers these parameters constant in time. We have found then that a nonzero $\sigma$, which is 
expected for any realistic fluid, typically escapes detection in this three-parameter model. It seems that it is enough to know $w$, and that $\clam$ 
and $\avis$ do not matter as long as they are positive.

The cosmological implication is that a canonical scalar field or perfect fluid representation might ultimately turn out to be a phenomenologically 
sufficient description of all the observational consequences of dark energy. The fundamental lesson is however that even then one cannot exclude, by 
appealing to observational evidence alone, the possibility of imperfectness in dark energy.

%

\section*{Acknowledgments} 

We thank Hans Kristian Eriksen for useful discussions. DFM  is supported by the Alexander von Humboldt Foundation. TK would like to 
thank the Magnus Ehrnrooth Foundation and the Finnish Cultural Foundation for support. JRK acknowledges support from the Research 
Council of Norway.






\bibliography{cites}

\end{document}